\documentclass[aps, prl,twocolumn,10pt,altaffilletter,nolongbibliography,numerical,flushbottom,secnumarabic,superscriptaddress,floatfix,nobibnotes]{revtex4-2}

\usepackage{graphicx}
\usepackage{braket}
\usepackage{amsmath, amssymb}
\usepackage{xfrac}
\usepackage{booktabs}  
\usepackage{comment}
\usepackage{physics}
\usepackage[normalem]{ulem}

\usepackage[breaklinks, pdftex, hyperfootnotes=true, pdfpagelabels, bookmarks, pageanchor]{hyperref}
\hypersetup{%
	colorlinks=true, linktocpage=true, pdfstartpage=1, pdfstartview=FitH, pdfborder={0 0 0},%
	breaklinks=true, pdfpagemode=UseNone, pageanchor=true, pdfpagemode=UseOutlines,%
	plainpages=false, bookmarksnumbered, bookmarksopen=true, bookmarksopenlevel=1,%
	hypertexnames=true, pdfhighlight=/O,
	urlcolor=blue, linkcolor=blue, citecolor=blue,
	}
	
\usepackage{etoolbox}	
\usepackage{orcidlink}

\newcommand\identity{1\kern-0.25em\text{l}}

\graphicspath{{figs/}}
 
\usepackage{soul}

\begin{document}

\title{Heat-to-motion conversion for quantum active matter}

\author{Alexander-Georg Penner$^{a}$}
\affiliation{Dahlem Center for Complex Quantum Systems and Fachbereich Physik, Freie Universit\"at Berlin, 14195 Berlin, Germany}
\thanks{These authors contributed equally to the work.}

\author{Ludmila Viotti$^{a}$}
\affiliation{The Abdus Salam International Center for Theoretical Physics,  34151 Trieste, Italy}
\thanks{These authors contributed equally to the work.}

\author{Rosario Fazio~\orcidlink{0000-0002-7793-179X}}
\affiliation{The Abdus Salam International Center for Theoretical Physics,  34151 Trieste, Italy}
\affiliation{Dipartimento di Fisica, Universit\`a di Napoli ``Federico II'', I-80126 Napoli, Italy}

\author{Liliana Arrachea~\orcidlink{0000-0002-7223-4610}}
\affiliation{Centro At\'omico Bariloche and Instituto de Nanociencia y Nanotecnolog\'ia CONICET-CNEA (8400), San Carlos de Bariloche, Argentina}

\author{Felix von Oppen~\orcidlink{0000-0002-2537-7256}}
\affiliation{Dahlem Center for Complex Quantum Systems and Fachbereich Physik, Freie Universit\"at Berlin, 14195 Berlin, Germany}

\date{\today} 

\begin{abstract}
We introduce a  model of an active quantum particle and discuss its properties. The  particle has a set of internal states  that mediate exchanges of heat with external reservoirs. Heat is then converted into motion by means of a spin-orbit term that couples internal and translational degrees of freedom. The quantum features of the active particle manifest both in the motion and in the heat-to-motion conversion. Furthermore, the stochastic nature of heat exchanges impacts the motion of the active particle and fluctuations can be orders of magnitude larger than the average values. The combination of spin-orbit interaction under nonequilibrium driving may bring active matter into the realm of cold atomic gases where our proposal can be implemented.  
\end{abstract}

\maketitle
\textit{Introduction ---} Active particles, of natural origin or artificially fabricated, propel themselves by converting environmental energy through nonequilibrium processes. The  field of active matter,  bridging between biology, chemistry, and physics has grown enormously over the years, leading to the discovery of a multitude of novel phenomena at the single- and many-particle levels. This body of work has been extensively described in several reviews and books, see for example~\cite{Marchetti2013hydrodynamics, Bechinger2016active,Ramaswamy2017active,Toner2024physics,sone2024hermitian}.  

While this effort has been overwhelmingly devoted to classical particles, it is natural to wonder whether active motion can also be realized in the quantum regime. Several new questions arise in this context. It would be important to understand necessary ingredients, identify suitable (possibly artificial) systems, and explore regimes realizing active quantum  particles. One would like to understand in which cases quantum effects enable,  influence, or enhance the functionalities of active particles, and explore  associated  signatures of genuine quantum behavior.
These questions naturally extend to the many-particle case of active matter, where new avenues of investigation emerge from collective quantum effects.  

Very recently, a few interesting works began to address some of these questions. A description of active quantum particles in terms of a non-Hermitian  quantum walk was put forward in~\cite{yamagishi2024proposal}, while in~\cite{zheng2024mimicking} the motion was induced by classical noise. Quantum active agents can eventually synchronize as discussed in~\cite{nadolny2024nonreciprocal}. Activity-induced collective effects were discussed in~\cite{adachi2022activity,khasseh2023active,yuan2024quantum,Takasan2024Apr} , 
employing both non-Hermitian~\cite{adachi2022activity,Takasan2024Apr}  and  Lindblad~\cite{khasseh2023active} dynamics. There are also pertinent connections to earlier works on directed motion and motors at the nanoscale \cite{Haenggi2009,Qi2009,Kudernac2011,Bustos2013,Meng2014,Arrachea2015,Bruch2018,vonLindenfels2019Aug}.

\begin{figure}
\includegraphics[width=.47\textwidth]{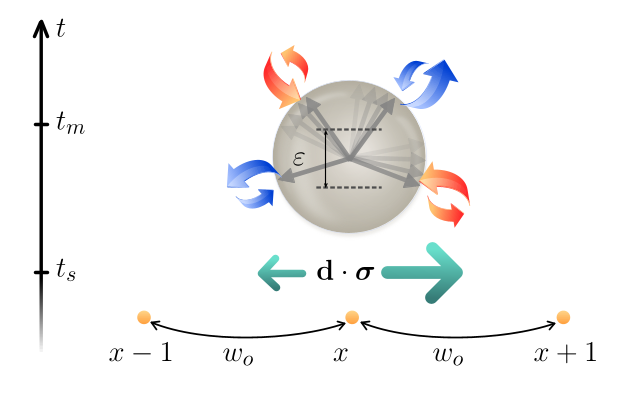} 
\caption{Illustration of the model for an active quantum particle. The particle is hopping with spin-dependent  amplitudes (black and green arrows) along a chain of sites. The spin of the particle is driven by spin-orbit ($\mathbf{d}$) and external magnetic ($\boldsymbol{\varepsilon}$) fields (gray arrows), and coupled to hot (red) and cold (blue) reservoirs. We assume that the spin relaxation time $t_s$ is shorter than the momentum relaxation time $t_m$ due to coupling to a phonon bath.}
\label{fig:sketch}
\end{figure}

In this work, we introduce a model of  active quantum particles, realizable with cold atoms, which establishes a connection to quantum heat engines~\cite{Kosloff2014quantum,cangemi2023quantum,Benenti2017Jun,Arrachea2023Jan}. The quantum particle  transforms the heat exchanged in the engine into motion by  coupling its internal and translational degrees of freedom. Quantum signatures  appear in the motion of the particle and play a crucial role in the heat-to-motion conversion. Furthermore, due to the microscopic nature of the quantum heat engine, fluctuations are relevant and the stochastic nature of the heat-to-motion conversion becomes important in the self-propulsion of the active particle. In this way, we provide a natural bridge between the fields of active matter and quantum thermodynamics~\cite{Vinjanampathy2016quantum,Goold2016role}. 

Our rather general approach applies to a large variety of specific cases. In the following, we will discuss the general idea and provide a detailed  analysis of a concrete example deep in the quantum limit: a spin-1/2 particle coupled to two heat baths. We will characterize its motion, its quantum signatures, and its stochastic nature. 

\textit{The model ---} As sketched in Fig.\ \ref{fig:sketch}, the  active quantum particle comprises internal and translational degrees of freedom. Its internal degrees of freedom (modeled as a set of discrete levels) couple to external reservoirs, which induce transitions between the levels. The exchange of energy with the environment fuels the motion of the particle when the internal and translational degrees of freedom are coupled, e.g., by spin-orbit interactions (see Ref.\ \cite{Lowe2018flocking} for a discussion of classical self-propelled particles inspired by spin-orbit coupling). Transitions between the internal levels give kicks to the particle. The rate at which the transitions take place depends on the coupling to and the state of the reservoirs, while the heat-to-motion conversion is controlled by the spin-orbit coupling. We also include weak coupling to a phonon bath which induces momentum relaxation. 

Directed motion of the quantum particle can be activated only under nonequilibrium conditions. For definiteness, we assume that nonequilibrium is achieved by keeping two reservoirs at different (hot/cold) temperatures $T_{h/c}= T \pm\Delta T/2$. Thus, the internal degrees of freedom of the particle constitute the working medium of a quantum heat engine fueling the motion. The coupling to these reservoirs is only through the internal degrees of freedom and does not select a preferred direction of motion. Nevertheless, nonequilibrium ensures that the emission and absorption of excitations from the different baths do not balance, allowing for self-propulsion of the particle. If $T_c=T_h=T$, no directed motion takes place. We emphasize that the choice of coupling the internal degrees of freedom to two reservoirs is only a specific example for operating under nonequilibrium conditions.

The quantum particle and its coupling to the environment are governed by the Hamiltonian  $\hat{{\cal H}} = \hat{{\cal H}}_{\rm p} + \hat{{\cal H}}_{\rm B} + \hat{V}_{\rm I}$.  The particle is described by  
\begin{equation}
    \hat{{\cal H}}_{\rm p} =\sum_{s, s^{\prime}}  \varepsilon_{s, s^{\prime}}\, |s\rangle\langle s^\prime | - \sum_{x,s, s^{\prime}} 
    w_{s,s^{\prime}} |x,s\rangle  \langle x+1,s^{\prime}|
    + {\rm h.c} \; ,
    \label{hsys}
\end{equation}
with $s$ labeling the set of discrete levels associated with the internal (``spin") degrees of freedom. The constants $\varepsilon_{s, s^{\prime}}$ account for the internal structure of the particle, while the $w_{s,s^{\prime}}$ describe spin-dependent hopping on a lattice of sites $x$, thus admitting processes in which hopping is accompanied by a change in the internal state of the particle (``spin-orbit" coupling). We consider a one-dimensional lattice, but the continuum limit or the extension to higher dimensions are straightforward. The bath Hamiltonian $\hat{{\cal H}}_{\rm B}$ describes sets of harmonic oscillators in thermal equilibrium at bath-dependent temperatures. The interaction between the baths and the active particle is described by the term $\hat{V}_{\rm I}$. Coupling to the spin degrees of freedom enables transitions between the internal states of the particle and therefore heat exchange between particle and thermal reservoirs. We further allow for processes affecting the translational degrees of freedom to account for momentum relaxation. Upon tracing over the environment, the state of the active quantum particle is described by a density matrix $\hat{\rho}$. 

The Hamiltonian $\mathcal{\hat H}$ embodies the essence of the heat-to-motion conversion and the connection between active quantum particles and quantum heat engines. We capture the main features of our approach within a minimal model: a quantum particle with two internal states ($s=\uparrow , \downarrow$). It is subject to a Zeeman field, $\varepsilon_{s,s'} = - [\boldsymbol{\varepsilon }\cdot \boldsymbol{\hat{\sigma}}]_{s,s'}$, and hopping amplitudes $w_{s,s'} = [w_o\, \hat{\identity}_s + i\,\mathbf{d}\cdot \boldsymbol{\hat{\sigma}}]_{s,s'}$ (with $\boldsymbol{\hat{\sigma}} = \left[\hat{\sigma}_x,\hat{\sigma}_y, \hat{\sigma}_z \right]$  the Pauli matrices and  $ \hat{\identity}_s$ the identity in spin space). The (effective) spin-orbit field $\mathbf{d}$ preserves time reversal symmetry when it is real, which we assume from now on. For the spin-$\sfrac{1}{2}$ particle, Eq.\ (\ref{hsys}) is diagonalized in momentum space by the composition of $k$-dependent rotations $\hat{R}_k$. Defining the rotated spin operators $\hat{\boldsymbol{\tau}}_{k} = \hat{R}_k^\dagger\, \hat{\boldsymbol{\sigma}}\, \hat{R}_k$, yields $\hat{{\cal H}}_{\rm p} = \sum_k \;(\epsilon_{k}\,\hat{\identity}_s + \Delta_k\, \hat{\tau}_{z,k})\ket{k}\!\bra{k}$, with $\epsilon_k \!=\! - 2\,w_o \cos k $, $\Delta_k \!=\! |\boldsymbol{\Delta}_k |$, $\boldsymbol{\Delta}_k  \!= \boldsymbol{\varepsilon} - 2\,\mathbf{d}_k$, and $\mathbf{d}_k = \mathbf{d}\sin k $.

We parametrize the density matrix of the $k$-block as $\hat{\rho}_k =r_{\! o,k}[ \, \hat{\identity}_s+ \mathbf{r}_k \cdot\boldsymbol{\hat{\tau}}_k]/2$, and assume that the system relaxes to thermal equilibrium at temperature $T=\beta^{-1}$ prior to applying the temperature bias. Tracing out the baths in the weak-coupling approximation leads to the Lindblad equation
\begin{equation}
    \dot{\hat{\rho}}_k =
     - i\, [\Delta_k \, \hat{\tau}_{z,k}\,,\,\hat{\rho}_k] + \mathcal{D}_\mathrm{s}[\hat{\rho}_k] + \mathcal{D}_\mathrm{m}[\{\hat{\rho}_{k'}\}],    \label{k-Lindblad} 
\end{equation}
where the dissipator 
\begin{equation}
      \mathcal{D}_\mathrm{s}[\hat{\rho}_k] =   \sum\limits_{\ell = \pm} \sum\limits_{\alpha=h,c} \Gamma^{(\ell )}_{\alpha,k} \left[\hat{\tau}^{\phantom{\dagger}}_{\ell,k} \hat{\rho}^{\phantom{\dagger}}_k \hat{\tau}^\dagger_{\ell,k} -
    \frac{1}{2}\{\hat{\tau}^\dagger_{\ell,k} \hat{\tau}^{\phantom{\dagger}}_{\ell,k}, \hat{\rho}^{\phantom{\dagger}}_k\} \right]
    \label{eq: spin dissipator}
\end{equation}
originates from the coupling ${\hat V}_I$ of the internal spin to the thermal baths \cite{sm_note}. Here, $\Gamma^{(\ell)}_{\alpha,k} =  \xi_{\alpha,k}\, [\,{\rm n}(2\Delta_k,T_{\alpha}) + \delta_{\ell, -}]$ with $\delta_{\ell, \pm}$ the Kronecker symbol and ${\rm n}(2\Delta_k,T_{\alpha})$ the Bose function at temperature $T_{\alpha}$. The factors $\xi_{\alpha,k} = \gamma_\alpha |\bra{\uparrow}\hat\tau_{\alpha,k}\ket{\downarrow}|^2$ allow for couplings to the two reservoirs with different strengths and through different spin operators. We denote $\hat{\tau}_{\pm, k} = (\hat{\tau}_{x,k} \pm i\, \hat{\tau}_{y,k})/2$, and $\left\{ \cdot,\cdot\right\}$ is the anticommutator. The dissipator $\mathcal{D}_m$ describes momentum relaxation as discussed further below. The dissipators $\mathcal{D}_s$ and $\mathcal{D}_m$  introduce two characteristic time scales, the spin and momentum relaxation times $t_s$ and $t_m$, respectively. We take momentum relaxation to be the longest time scale in the problem, $t_m\gg t_s$ (see Fig.\ \ref{fig:sketch}). Since $\mathcal{D}_s$ is diagonal in momentum, the momentum distribution $r_{o,k}$ remains constant for times short compared to $t_m$, even in the presence of a temperature imbalance $\Delta T$. In contrast, the temperature bias causes  nontrivial dynamics of the state of the internal degrees of freedom, as  encoded in $\mathbf{r}_k(t)$. In particular, this dynamics -- and hence the dynamics of the active particle -- are genuinely quantum due to the  $k$-dependent quantization axis of the spin implicit in Eq.\ (\ref{k-Lindblad}).

\begin{figure}
\includegraphics[width=\linewidth]{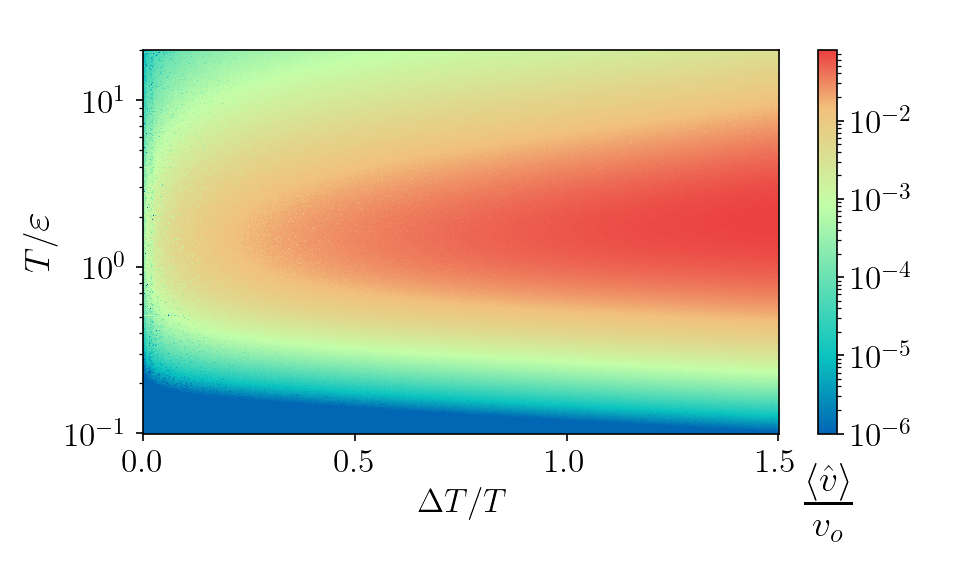} 
\caption{Average velocity of the particle for $t_s \ll t \ll t_m$ {\em{vs}}.\ equilibrium temperature and (relative) temperature bias $\Delta T/T$. Velocity scale: $v_o = \varepsilon\,a$.  Parameters: Zeeman field $\boldsymbol{\varepsilon} = \varepsilon\,(\hat{x} \cos \theta + \hat{z} \sin \theta)$, with $\theta = 3\pi/4 - \delta$ and $\delta = 0.01$, hopping energy $w_o = 2\,\varepsilon$, spin-orbit coupling $\mathbf{d}/\varepsilon =\sfrac{0.6}{\sqrt{2}}\, (\hat{x} + \hat{z})$. Particle couples to baths through $\hat\tau_{h,k}\! =\! \hat\tau_{x,k}$ and $\hat\tau_{c, k}\! =\! \hat\tau_{z,k}$ with $\gamma= 0.2\,\varepsilon$. } 
\label{fig:velocity}
\end{figure}

\textit{Heat-to-motion conversion ---} We use Eq.\ (\ref{k-Lindblad}) to analyze the dynamics of the active quantum particle. Its average velocity is defined as $
\langle \hat{v}(t) \rangle  = \sum_k \mbox{Tr}_s \{ \hat{\rho}_k(t)\, {\hat v}_k \} $, where $\hat v_k = \partial_k \epsilon_k  + (\partial_k \hat {\boldsymbol{\Delta}}_k) \cdot {\hat {\boldsymbol{\sigma}}}$ is the velocity operator. It vanishes in equilibrium, as can be verified by substituting the Boltzmann distribution. Under nonequilibrium conditions, the nano-engine will start to fuel the particle through heat-to-motion conversion. In linear response to a small temperature difference, and in the absence of momentum relaxation (i.e., for times $t_s\ll t \ll t_m$), we find the average steady-state velocity 
\begin{equation}
    \langle \hat{v}\rangle 
    =
    \frac{\Delta T}{T^2}\sum_kr_{o,k}\frac{\xi_{h,k}-\xi_{c,k}}{\xi_{h,k}+\xi_{c,k}} \frac{\boldsymbol{\Delta}_k \cdot \partial_k\mathbf{d}_k}{ \cosh^2(\beta \Delta_k)} \; .
\label{eq:vel}
\end{equation}
Here, the momentum distribution $r_{o,k}$ is thermal,
\begin{equation}
r_{o,k}^\mathrm{th} =\frac{1}{\mathcal{Z}} e^{- \beta \epsilon_k} \cosh(\beta \Delta_k),
\label{eq:ro_thermal}
\end{equation}
where $\mathcal{Z} = \mbox{Tr}\, e^{-\beta\, \hat{\cal H}_{\rm p}}$ denotes the equilibrium partition function of the particle. 
In our case with only two levels, the velocity is nonzero when the particle couples to the reservoirs with different amplitudes $\xi_{h,k} \neq \xi_{c,k}$. In numerical calculations, we implement this difference by coupling to the baths through different Pauli operators, while keeping equal strengths, $\gamma_h = \gamma_c =\gamma$. 
Equation \eqref{eq:vel} implies that a nonzero average velocity  requires that $w_0$, $\boldsymbol{\epsilon}$, and $\mathbf{d}$ are all nonvanishing \cite{sm_note}.

Figure \ref{fig:velocity} shows the steady-state velocity as a function of temperature bias $\Delta T/T$ and temperature $T$ in the absence of momentum relaxation. The steady-state velocity has an interesting $T$ dependence, vanishing in both the high and low-temperature limits with an intermediate optimal temperature. At low temperatures, the average velocity vanishes exponentially, $ \langle \hat{v} \rangle  \sim e^{-2\beta \varepsilon}$, while at high temperatures it goes to zero as $\langle \hat{v} \rangle  \sim 1/T^3$. This can be understood by noting that at low temperatures, transitions are suppressed and therefore the quantum engine is unable to provide the necessary imbalance to activate the particle. With increasing temperature, the relative difference in occupations between the two reservoirs decreases and the deviations from equilibrium are less pronounced. This behavior persists beyond linear response as shown in Fig.\ \ref{fig:velocity}. For $w_o\gg d,\varepsilon$, Eq.\ \eqref{eq:vel} gives an optimal temperature of $0.83\varepsilon$ \cite{sm_note}.

\begin{figure}
    \centering
    \includegraphics[width=\linewidth]{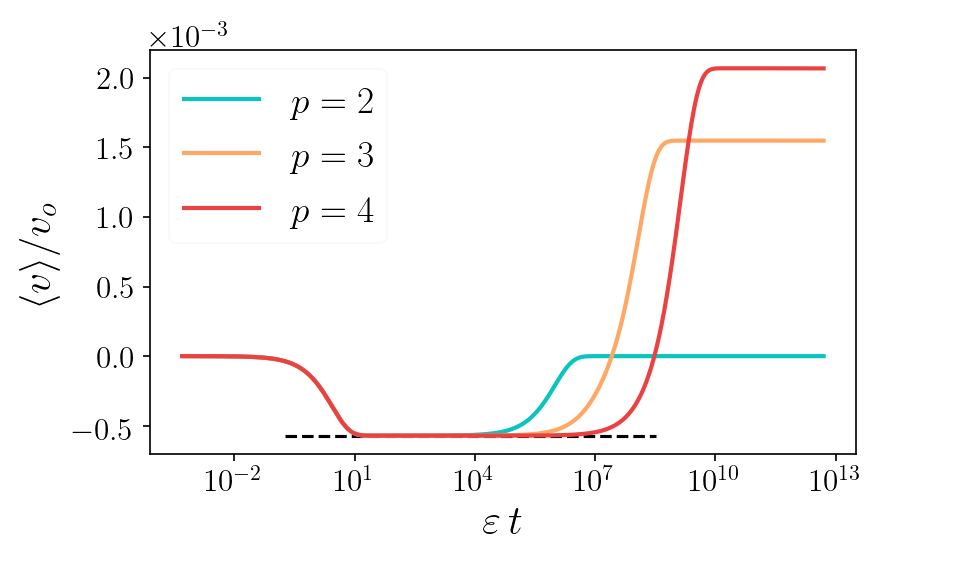}

    \caption{Average velocity in Eq.\ (\ref{eq:vel}) {\em{vs}}.\ time, contrasting states with and without momentum relaxation. State without momentum relaxation is indicated by black dashed line. We consider phonon baths with various densities of states $\nu\sim |\omega|^p$ (see legend). Parameters: same as in Fig.\  \ref{fig:velocity} with bath temperatures $T = \varepsilon$ and $\Delta T/T = 0.01$, and momentum coupling decay defined by $q_o = 0.02\,\varepsilon$. Hierarchy of time scales $t_s\ll t_m$  guaranteed by $\lambda_0/N =0.002\,\varepsilon\ll\gamma$}
    \label{fig:vt}
\end{figure}

\textit{Momentum relaxation ---} In general, the complex environment of the particle induces additional processes, which randomize its momentum. This can be modeled by introducing a phonon bath at temperature $T$ with coupling Hamiltonian
\begin{equation}
    H_r = \sum_{k,q,n} \sqrt{\frac{\lambda_q}{N}}\, |k\rangle \langle k+q| \otimes \hat{\identity}_s \otimes (b_{-q,n} + b^\dagger_{q,n}).
\label{phonon-ham}
\end{equation} 
Here, $N$ denotes the number of sites and $b_{-q,n}$ ($b^\dagger_{q,n}$) annihilates (creates) a phonon in transverse mode $n$ with longitudinal momentum $q$. Due to the $k$-dependent rotations $R_k$, momentum-relaxation processes may or may not flip the spin. We show in \cite{sm_note} that spin-conserving transitions dominate as long as the coupling $\lambda_q$ decays rapidly with $q$ on a scale $q_0$, which is the smallest momentum scale in the problem, and the phonon density of states $\nu \sim |\omega|^{p}$ does not grow too rapidly with $\omega$, namely, for $p=1,2,3$. We specify to this case in the following. 

Tracing out the phonon bath introduces the dissipator $\mathcal{D}_m$ into Eq.\ (\ref{k-Lindblad}). We relegate explicit expressions and calculational details to  \cite{sm_note}. We find that the equation for $\mathbf{r}_k$ decouples from the momentum distribution $r_{o,k}$, so that the spin distribution $\mathbf{r}_k$ is independent of the momentum distribution $r_{o,k}$. As transitions are local in momentum space, the latter obeys a drift-diffusion equation, 
\begin{equation}
    \partial_t r_{o,k} = \partial_k[v_D(k)r_{o,k} + D(k)\partial_k r_{o,k}].
    \label{eq:ddd}
\end{equation}
The drift velocity $v_D(k)$ and the diffusion constant $D(k)$ depend on $k$ and on the spin distribution $\mathbf{r}_{k}$. The steady-state velocity in the presence of momentum relaxation is still given by Eq.\ \eqref{eq:vel}, albeit with $r_{o,k}$ modified to be the stationary solution $r_{o,k}^\infty$ of the drift-diffusion equation. For $p=2$, we find 
\begin{equation}
    r_{o,k}^\infty = \frac{1}{\mathcal{Z}}\exp{-\frac{\beta}{2}\int_0^k \dd k' \Tr_s\left(\hat{v}_{k'}[ \, \hat{\identity}_s+ \mathbf{r}_{k'} \cdot\boldsymbol{\sigma}]\right)}.
\end{equation}
We give results for other values of $p$ in Ref.\ \cite{sm_note}.

We use numerical simulations to illustrate the evolution of the average velocity of the active particle across the entire range of time scales in Fig.\ \ref{fig:vt}, contrasting various  densities of states of the phonon bath. At times short compared to $t_m$, the velocity is independent of the phonon bath. After an initial transient ($t_s\ll t \ll t_m$), the active particle moves at the steady-state velocity in the absence of momentum relaxation (see Eq.\ \eqref{eq:vel}, indicated by the dashed line in the figure) by converting the heat exchanged with the baths into motion. At late times compared to the momentum relaxation time, the average velocity reaches a steady state, which is distinct from the steady-state velocity of the model without momentum relaxation. In fact, the two steady-state velocities can differ not only in magnitude, but also in sign. Remarkably, as seen in Fig.\ \ref{fig:vt}, momentum relaxation can induce an increase in the magnitude of the average velocity. We also observe that once phonon relaxation sets in, the average velocity depends sensitively on the phonon density of states $\nu$. The time dependence remains qualitatively unchanged when momentum-relaxation processes with spin flips become important ($p=4$).  

\begin{figure}
\includegraphics[width=.45\textwidth]{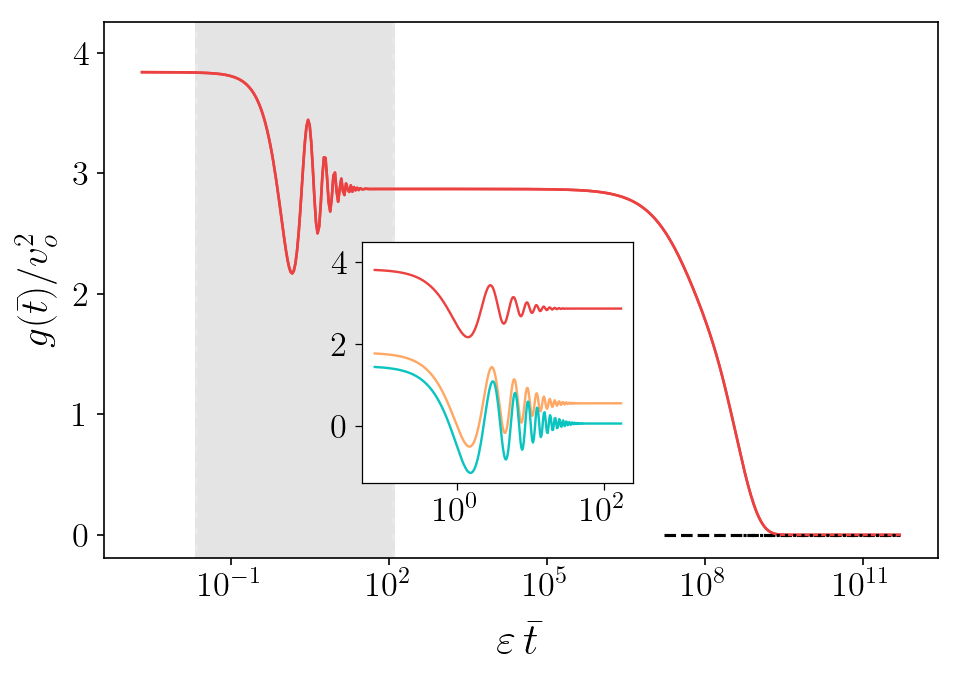} 
\caption{Velocity correlator for $t \gg t_m$ {\em{vs}}.\ relative time $\bar t$. Parameters: Same as Fig.\ \ref{fig:vt}, with $p = 3$.
Inset: zoom of oscillations observed for $\bar t\ll t_m$ (gray background in main plot) for various equilibrium temperatures:  $T = 0.02\,\varepsilon$ (green), $T = 0.2\,\varepsilon$ (orange), and $T = \varepsilon$ (red) with  temperature bias $\Delta T = 0.01\,T$. 
}
\label{fig:correlator}
\end{figure}

\textit{Quantum effects and velocity correlations ---} The quantum nature of the active particle becomes apparent in the velocity-velocity correlation function 
\begin{equation}
    g(t, \bar t) = \langle  \delta \hat{v}(t) \, \delta\hat{v}(t+\bar t)\rangle,
\end{equation}
where $\delta \hat{v}(t)  = \hat{v}(t) - \langle \hat{v}(t) \rangle $. We compute the steady-state correlation function using the quantum regression theorem~\cite{Carmichael1993open}, 
\begin{align} 
    g(\bar t)  = \sum_k r_{o,k}\left( e^{-\Gamma_k \bar{t}/2}\cos(2\Delta_k \bar{t})(\hat{\boldsymbol{\Delta}}_k \times \partial_k \mathbf{d}_k)^2\right. \nonumber \\
    +\,4\,e^{-\Gamma_k \bar{t}}\sum_{\alpha,\alpha'} \frac{\Gamma^+_{\alpha,k}\Gamma^-_{\alpha',k}}{\Gamma_k}\left.(\hat{\boldsymbol{\Delta}}_k \cdot \partial_k \mathbf{d}_k)^2\right) + g_o,
    \label{eq:correlator}
\end{align}
where $\Gamma_k = \sum_{\ell , \alpha}\Gamma^{(\ell )}_{\alpha,k}$ (see \cite{sm_note} for details) and $\bar t\ll t_m$. When evaluated with the appropriate momentum distribution, this expression applies to the steady states with and without momentum relaxation. The constant $g_o$ reflects residual correlations for  $t_s \ll \bar{t}\ll t_m$. These correlations decay in the long-time limit $t,\bar{t} \gg t_m$, where $g(t,\bar{t})\to 0$. The velocity fluctuations far exceed the average velocity, emphasizing the importance of stochasticity.

Figure \ref{fig:correlator} shows a typical time trace of the velocity correlations for $t\gg t_m$. The oscillations directly reflect the Rabi oscillations of the internal degrees of  freedom in the motion of the particle. These exist in equilibrium and remain clearly visible for $\Delta T \ne 0$, when the quantum particle is active. The oscillation period and temperature scale are controlled by the Zeeman field, while the oscillations decay on the time scale $t_s$. These quantum features are sensitive to the relative orientation of the vectors $\mathbf{d}$ and $\boldsymbol{\Delta}_k$, see Eq.\ \eqref{eq:correlator}. For $\mathbf{d}\parallel \boldsymbol{\Delta}_k$, energy is exchanged through incoherent transitions between the two internal states and the oscillations disappear. 

\textit{Conclusions ---} We set up a framework to engineer active particles in the quantum regime. Its essential ingredients are the presence of ``spin-orbit" coupling between internal and translational degrees of freedom, breaking of time-reversal symmetry by a ``magnetic'' field, and an imbalance in the couplings to the energy reservoirs, which drive the system out of equilibrium. These ingredients are available both in cold-atom and trapped-ion systems. In the extreme quantum limit of a particle with two internal states, we explored the properties of the heat-to-motion conversion, the quantum signatures of the active motion, and the consequences of a noisy environment leading to momentum relaxation. More generally, we expect the qualitative results to extend beyond the specific model considered, making our proposed approach quite amenable to experimental implementations.  
 
So far, we focused on single-particle properties in one dimension. The model can be straightforwardly extended to include interactions between  many particles as well as higher dimensional systems. This opens the path to a wealth of additional phenomena such as collective behavior, topological phenomena, etc. More broadly, the dynamics of these active particles bear features typical of quantum walks~\cite{Ambainis2001Jul,Dur2002Nov,peruzzo2010quantum} and are closely related to quantum thermodynamics~\cite{Vinjanampathy2016quantum,Goold2016role}, which opens appealing bridges to other classes of phenomena.

\begin{acknowledgments}
{\em Acknowledgments.---}We  thank M.\ Brunelli, A.\ Jelic, C.\  Schmiegelow, and M.\ Schir\`o for fruitful discussions. We acknowledge financial support by PNRR MUR project PE0000023- NQSTI (R.F. and L.V.), by the European Union (ERC - RAVE, 101053159) (R.F.), by CONICET, PICT 2020-A-03661 Argentina, ICTP Trieste, IHP (UAR 839 CNRS-Sorbonne Universit\'e), the LabEx CARMIN (ANR-10-LABX-59-01), the Alexander-von-Humboldt foundation (L.A.), and by CRC 183 of Deutsche Forschungsgemeinschaft as well (FvO). Views and opinions expressed are those of the authors only and do not necessarily reflect those of the European Union or the European Research Council. Neither the European  Union nor the granting authority can be held responsible for them.
\end{acknowledgments}


%

\onecolumngrid

\clearpage

\setcounter{figure}{0}
\setcounter{section}{0}
\setcounter{equation}{0}
\renewcommand{\theequation}{S\arabic{equation}}
\renewcommand{\thefigure}{S\arabic{figure}}

\onecolumngrid

\centerline{\textbf{\large Supplemental Material: Heat-to-motion conversion for quantum active matter}}
\medskip

\centerline{Alexander-Georg Penner, Ludmila Viotti, Rosario Fazio, Liliana Arrachea, Felix von Oppen}

\section{Average velocity} 

We provide details of the calculation of the active particle's average velocity in the intermediate steady state for times $t_s\ll t \ll t_m$. We derive evolution equations for the parameters $r_{o,k}$ and $\mathbf{r}_k$ of $\hat{\rho}_k$ by taking the trace after multiplying Eq.\ (\ref{k-Lindblad}) of the main text by a unit matrix or the vector of Pauli matrices ${\boldsymbol{\tau}}_{k}$,
\begin{align}
    &\partial_t r_{o,k} = 0, \ \ \ \partial_t r_{x,k} = -2\Delta_kr_{y,k} -\frac{1}{2}\sum\limits_{\ell = \pm} \sum\limits_{\alpha=h,c} \Gamma^{(\ell )}_{\alpha,k} r_{x,k} , \nonumber\\
    &\partial_t r_{y,k} = 2\Delta_kr_{x,k} -\frac{1}{2}\sum\limits_{\ell = \pm} \sum\limits_{\alpha=h,c} \Gamma^{(\ell )}_{\alpha,k} r_{y,k} , \ \ \ \partial_t r_{z,k} = \sum_{\alpha = h,c}(\Gamma^-_{\alpha,k}-\Gamma^+_{\alpha,k}) - \sum\limits_{\ell = \pm} \sum\limits_{\alpha=h,c}\Gamma^{(\ell)}_{\alpha,k} r_{z,k}. 
    \label{parameter evolution equations}
\end{align}
Notice that the coupling to the spin baths leaves the momentum distribution $r_{o,k}$ unchanged. Moreover, the time evolution of the spin distribution $\mathbf{r}_k$ decouples from $r_{o,k}$. 

From Eq.\ (\ref{parameter evolution equations}), we deduce the steady state $\hat{\rho}_{k} = r_{o,k}(1+r_{z,k}\tau_{z,k})/2$ with 
\begin{equation}
    r_{o,k} = \frac{1}{\mathcal{Z}} e^{-\beta \epsilon_k }\cosh(\beta \Delta_k),\ \ \  r_{z,k} =  \sum\limits_{\alpha=h,c} \xi_{\alpha,k}\left[\sum\limits_{\alpha=h,c} \xi_{\alpha,k}\left\{ \rm{n}(2\Delta_k,T_\alpha)+1\right\}\right]^{-1}.
    \label{static state}
\end{equation}
The expression for $r_{o,k}$ follows from the assumption that the system is initially in thermal equilibrium with $\Delta T=0$. Expanding $r_{z,k}$ to linear order in $\Delta T $ yields 
\begin{equation}
    r_{z,k} = \tanh(\beta \Delta_k) - \Delta T \frac{\Delta_k}{2T^2}\frac{\xi_{h,k}-\xi_{c,k}}{\xi_{h,k}+\xi_{c,k}}\frac{1}{\cosh^2(\beta \Delta_k)} + \mathcal{O}(\Delta T^2).
\end{equation}
As discussed in the main text, the zeroth-order term in $\Delta T$ gives a vanishing contribution to the average velocity. To linear order in $\Delta T$, we find \begin{equation}
    \langle \hat{v}\rangle 
    =
    \frac{1}{\mathcal{Z}}\sum_k\frac{\xi_{h,k}-\xi_{c,k}}{\xi_{h,k}+\xi_{c,k}} e^{-\beta \epsilon_k} \frac{\boldsymbol{\Delta}_k \cdot \partial_k\mathbf{d}_k}{ \cosh(\beta \Delta_k)} \; \frac{\Delta T}{T^2} +\mathcal{O}(\Delta T^2).
    \label{average velocity}
\end{equation}
for the average velocity of the active particle.

The average velocity vanishes, when any of the parameters $\mathbf{d}$, $\boldsymbol{\varepsilon}$, or $w_o$ vanish. This is explicit in Eq.\ \eqref{average velocity} for $\mathbf{d}=0$. For $\boldsymbol{\varepsilon}=0$, we observe that the $\xi_{\alpha,k}$ become $k$-independent, while the remaining integrand is odd in $k$. Finally, for $w_o=0$, the integral over $k$ vanishes as it takes the form $\sum_k f(\sin k) \cos k$. 

The expression for the average velocity can be evaluated further in a few temperature regimes. We assume that the band width $w_o$ is large compared to the Zeeman and spin-orbit field. In the low-$T$ limit, $T \ll \Delta_k\ll w_o$, the sum over $k$ in Eq. (\ref{average velocity}) is dominated by its value at $k=0$. This yields
\begin{equation}
    \langle \hat{v} \rangle \simeq \frac{\xi_{h,0}-\xi_{c,0}}{\xi_{h,0}+\xi_{c,0}}\frac{\boldsymbol{\Delta}_0\cdot \partial_k \mathbf{d}_k|_{k=0}}{\cosh^2(\beta \Delta_0)}\frac{\Delta T}{T^2} = \frac{\xi_{h,0}-\xi_{c,0}}{\xi_{h,0}+\xi_{c,0}} 4 \boldsymbol{\varepsilon}\cdot  \mathbf{d} e^{-2\beta \varepsilon} \frac{\Delta T}{T^2} . 
\end{equation} 
For intermediate temperatures, $\Delta_k, T\ll w_o$, $k=0$ still dominates and we find 
\begin{equation}
    \langle v \rangle \simeq\frac{\xi_{h,0}-\xi_{c,0}}{\xi_{h,0}+\xi_{c,0}}\frac{\boldsymbol{\varepsilon}\cdot  \mathbf{d}}{\cosh^2(\beta \varepsilon)}\frac{\Delta T}{T^2}.
    \label{intermediate}
\end{equation}
We can extract the optimal temperature $T_{\text{opt}}$, for which the average velocity is maximal. Differentiating Eq.\ \eqref{intermediate} with respect to $T$ yields the condition
\begin{equation}
    T_{\text{opt}} = \varepsilon \tanh(\frac{\varepsilon}{T_{\text{opt}}}).
\end{equation}
This gives $T_{\text{opt}} \simeq 0.83\varepsilon.$

Finally, in the high temperature limit, $\Delta_k \ll w_o \ll T$, we expand $e^{-\beta \epsilon_k}$ in powers of $\beta$. The zeroth-order term vanishes as it is of the form $\sum_k f(\mathbf{d}_k) \partial_k \mathbf{d}_k$ with $\mathbf{d}_k$ a periodic function with zero mean. In linear order, we find  
\begin{equation}
    \langle v \rangle \simeq \frac{2}{\mathcal{Z}}\sum_k\frac{\xi_{h,k}-\xi_{c,k}}{\xi_{h,k}+\xi_{c,k}}  \; w_o\cos k \;\boldsymbol{\Delta}_k \cdot \partial_k\mathbf{d}_k \frac{\Delta T}{T^3}.
\end{equation}
Simplifying to $\boldsymbol{\varepsilon} \parallel \mathbf{d}$, the factors $\xi_{\alpha,k}$ no longer depend on $k$. Evaluating the sum explicitly gives
\begin{equation}
    \langle v \rangle \simeq \frac{\xi_h - \xi_c}{\xi_h + \xi_c} w_o \boldsymbol{\varepsilon}\cdot  \mathbf{d} \frac{\Delta T}{T^3}.
\end{equation}
In particular, these expressions agree with the temperature dependencies mentioned in the main text. 

\section{Momentum relaxation} 

\begin{figure}
    \centering    \includegraphics[width=0.45\linewidth]{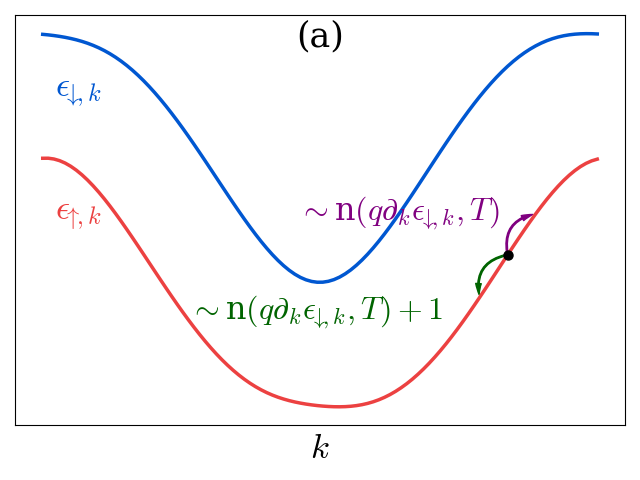}
    \includegraphics[width=0.45\linewidth]{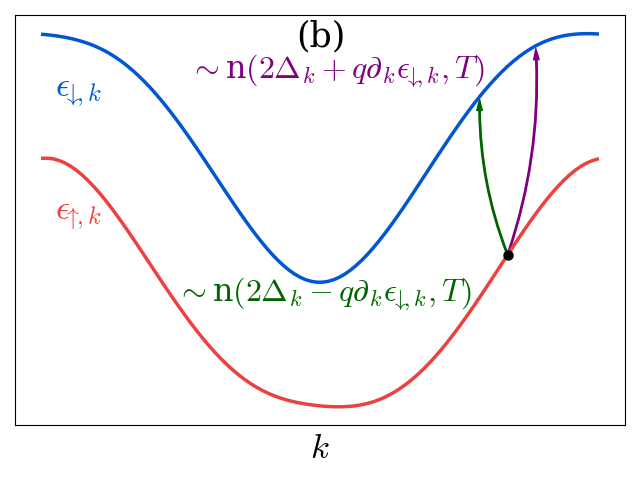}
    \caption{Momentum relaxation processes. (a) Spin-conserving scattering processes between momenta $k$ and $k \pm q$ with $q$ small (of order $q_0$). (b) Spin-flip scattering processes between momenta $k$ and $k \pm q$.  The dispersions $\epsilon_{\uparrow,k}$ and $\epsilon_{\downarrow,k}$ are calculated for the parameters used in Fig.\ 2 of the main text. }
    \label{fig: momentum relaxation processes}
\end{figure}

The coupling to the phonon bath in Eq.\ (\ref{phonon-ham}) (main text) leads to momentum relaxation. We choose a coupling
\begin{equation}
    \lambda_q = \lambda_0 \exp(-\frac{|q|}{q_0})
\end{equation}
with a small $q_0$. Thus, phonon processes are restricted to scattering between closeby momentum states. This allows for deriving a local evolution equation for the momentum distribution. As the overall probability is conserved, this takes the form of a drift-diffusion equation, see Eq.\ \eqref{eq:ddd} (main text). We give details of the derivation in the remainder of this section.

Momentum-relaxation processes may or may not change the spin state of the particle due to the $k$-dependence of the effective Zeeman field. Provided that the phonon density of states $\nu$ does not increase too rapidly at small energies, we find that momentum relaxation is dominated by spin-conserving processes. This can be seen by extracting the scaling of the drift velocity and the diffusion constant with $q_0$. The underlying momentum relaxation processes are illustrated in Fig.\ \ref{fig: momentum relaxation processes}. Panel (a) shows spin-conserving processes. Here, scattering between nearby momentum states is due to phonons with energy of order $\omega \sim q$ (with $q$ of order  $q_0$). The corresponding scattering rates are determined by the phonon density of states $\nu(\omega)$ as well as the Bose distribution $\text{n}(\omega,T)$ with $\omega = \epsilon_{\sigma,k\pm q}-\epsilon_{\sigma,k} \simeq \pm q \partial_k \epsilon_{\sigma,k}$. Panel (b) shows the corresponding processes for spin-flip scattering. Due to the spin splitting, the relevant phonon energies approach a nonzero constant for $q \to 0$, i.e,. $\omega \simeq 2\Delta_k \pm q \partial_k \epsilon_{\sigma,k}$. We estimate the drift velocity $v_D(k)$ and the diffusion coefficient $D(k)$ based on the scattering rates, assuming a phonon density of states $\nu(\omega) \sim |\omega|^p$. We can estimate the diffusion constant (drift velocity) through the average (difference) of the scattering rates to the left and right, multiplied by $q^2$ ($q$). (Due to the $k$-dependence of the rates, there are additional contributions. However, these do not modify the conclusions.) The contributions of spin-conserving scattering can thus be estimated as 
\begin{align}
    v^{(a)}_D(k) &\sim q \frac{\nu(\omega)}{|\omega|} \frac{\text{n}(\omega,T)+1-\text{n}(\omega,T)}{2} \sim \frac{1}{2}q^{p} |\partial_k \epsilon_{\sigma,k}|^{p-1}, \nonumber \\
    D^{(a)}(k) &\sim q^2 \frac{\nu(\omega)}{|\omega|} \frac{2\text{n}(\omega,T)+1}{2} \sim  q^{p} T|\partial_k \epsilon_{\sigma,k}|^{p-2}.
    \label{eq: heuristic vD}
\end{align}
Here, we used that $T\gg \omega \sim q$. The factors of $1/|\omega|$ emerge, because momentum along the chain is conserved in the scattering process. The contribution of the spin-flip processes differ in that the relevant phonon energies do not vanish for $q\to 0$, implying that the density of states as well as the Bose function can be approximated by a $q$-independent constant. Moreover, the matrix element for spin-flip scattering is proportional to $q$. Thus, the contributions of spin-flip scattering scales with $q$ as 
\begin{align}
    v^{(b)}_D(k) &\sim  q^{4}  \partial_{k} \epsilon_{\sigma,k}, \nonumber\\
    D^{(b)}(k) &\sim  q^{4}  .
\end{align}
We can now compare the contributions of spin-conserving and spin-flip scattering. As we consider $q_0$ as the smallest momentum scale, we need to retain only the contribution of the spin-conserving scattering processes, provided that $p=1,2,3$, which we assume in the following. (We note that $\nu(\omega)\sim |\omega|^{d-1}$ for a $d$-dimensional bath of acoustic phonons.)

In view of these estimates, we only retain the contribution of spin-conserving scattering in the dissipator due to the momentum bath, 
\begin{equation}
    \mathcal{D}_m[\hat{\rho}] = \sum_{k,q,\tau,\tau'} \frac{\lambda_q}{N}\nu_\perp(\Omega^{k,q}_{\tau,\tau'},q)\left|\braket{\tau_k}{\tau'_{k+q}}\right|^2|\rm{n}(\Omega^{k,q}_{\tau,\tau'},T)|\left(\ketbra{k + q,\tau'}{k,\tau} \hat{\rho} \ketbra{k,\tau}{k +q,\tau'}-\frac{1}{2}\{\ketbra{k,\tau}{k,\tau},\hat{\rho}\}\right),
    \label{eq:DmS}
\end{equation}
where $\tau$ labels the eigenstates of $\tau_{z,k}$ and we use the shorthand  $\Omega^{k,q}_{\tau,\tau'} = \epsilon_{\tau',k+q} - \epsilon_{\tau,k} $. Moreover, we define
\begin{equation}
    \nu_\perp(\Omega,q) = 2\pi \sum_{\mathbf{q}_\perp} \delta(|\Omega| - \omega_\mathbf{q})
\end{equation}
with $\omega_\mathbf{q}$ the dispersion of the phonon bath. For $\omega\sim q$ as in Eq.\ \eqref{eq:DmS}, one has $\nu_\perp(\omega)\sim |\omega|^{p-1}$ for acoustic phonons, if the full phonon density of states is $\nu(\omega,q)\sim |\omega|^p$. We note that the prefactor in $\nu_\perp(\omega,q)$ is independent of $k$, provided that the phonon velocity is small compared to typical particle velocities $\partial_k\epsilon_{\tau,k}$. Explicitly, we find $\nu_\perp(\omega,q)= A|\omega|^{p-1}/c^p$ with phonon velocity $c$ and a numerical prefactor $A$.

We trace the evolution equation $\partial_t \hat{\rho} = \mathcal{D}_m[\hat{\rho}]$ over spin space (we suppress $\mathcal{D}_s$ as it does not contribute to the evolution of the momentum distribution),

\begin{equation}
   \partial_t r_{o,k} = \sum_{k,q,\tau,\tau'} \frac{\lambda_q}{N}\nu_\perp(\Omega^{k,q}_{\tau,\tau'},q)\left|\braket{\tau_k}{\tau'_{k+q}}\right|^2(|\text{n}(\Omega^{k,q}_{\tau,\tau'},T)+1|r_{o,k+q,\tau} - |\text{n}(\Omega^{k,q}_{\tau,\tau'},T)|r_{o,k,\tau}),
\end{equation}
where $r_{o,k,\tau}$ denotes the momentum distribution for spin $\tau$ defined through
\begin{equation}
    r_{o,k,\uparrow} = r_{o,k}(1+r_{z,k})/2, \ \ \ r_{o,k,\downarrow} = r_{o,k}(1-r_{z,k})/2.
\end{equation}
Taking the continuum limit, we transform the rate equation into a drift-diffusion equation,
\begin{equation}
    \partial_t r_{o,k} = \partial_k \left[v_D(k)r_{o,k} + D(k)\partial_k r_{o,k}\right],
\end{equation}
for the momentum distribution. Working to leading order in powers of $q_0$ and keeping only spin-conserving scattering processes, the drift velocity becomes (taking $a=1$ for the lattice constant)
\begin{align}
    v_D(k) &= \frac{A}{c^{p}}\left[\int_0^\infty \frac{\dd q}{2\pi} \lambda_q q^{p}\right]\left[(1+r_{z,k})(\partial_k \epsilon_{\uparrow,k})^{p-1}\text{sgn}(\partial_k \epsilon_{\uparrow,k})^{p} + (1-r_{z,k})(\partial_k \epsilon_{\downarrow,k})^{p-1}\text{sgn}(\partial_k \epsilon_{\downarrow,k})^{p}\right. \nonumber\\ & \qquad \qquad \qquad +\left. T \partial_k r_{z,k}(|\partial_k \epsilon_{\uparrow,k}|^{p-2}-|\partial_k \epsilon_{\downarrow,k}|^{p-2})\right]
\end{align}
and the diffusion coefficient takes the form
\begin{equation}
    D(k) = \frac{AT}{c^{p}}\left[\int_0^\infty \frac{\dd q}{2\pi} \lambda_q q^{p}\right][(1+r_{z,k})|\partial_k \epsilon_{\uparrow,k}|^{p-2} + (1-r_{z,k})|\partial_k \epsilon_{\downarrow,k}|^{p-2}].
\end{equation}These expressions are consistent with the estimates in Eq.\ (\ref{eq: heuristic vD}). We note that the parameters entering the drift-diffusion equation depend on the spin distribution $r_{z,k}$. Moreover, the momentum relaxation processes also contribute to the evolution equation of $r_{z,k}$. These originate either in the subdominant spin-flip processes or the difference between the spin-conserving scattering rates for the spin-up and spin-down components. However, we assume that momentum relaxation is much slower than the processes due to the spin baths, so that this can be neglected. This also implies that $r_{z,k}$ in $v_D(k)$ and $D(k)$ is independent of time. 

The steady-state solution of the drift-diffusion equation gives the momentum distribution
\begin{equation}
    r_{o,k} = \frac{1}{\mathcal{Z}}\exp{-\int_{0}^k \dd k' \frac{v_D(k')}{D(k')}}.
\end{equation}
For $p=2$, the steady state of the momentum occupation numbers may also be written as 
\begin{equation}
    r_{o,k} = \frac{1}{\mathcal{Z}} \exp{-\frac{\beta}{2}\int_0^k \dd k' \Tr_s(\hat{v}_{k'}(\identity_s + r_{z,k'}\hat{\boldsymbol{\Delta}}_{k'}\cdot \boldsymbol{\sigma}))}
\end{equation}
The average steady-state velocity is given by
\begin{equation}
    \langle \hat{v} \rangle = \sum_k \Tr_s (\hat{v}_k \hat{\rho}_k) = \sum_k \frac{r_{o,k}}{2} \Tr_s(\hat{v}_{k}(\identity_s + r_{z,k}\hat{\boldsymbol{\Delta}}_k\cdot \boldsymbol{\sigma})) = -T \sum_k \partial_k r_{o,k},
\end{equation}
and vanishes to lowest order in $q_0$. Importantly, this result for $p=2$ does not carry over to other dependencies of the phonon density of states such as $p = 3$, for which one obtains a nonzero average velocity in the momentum relaxed state. 

We can use the drift-diffusion equation to estimate the momentum relaxation time $t_m$. An estimate of $t_m$ is given by the time it takes to diffuse across the relevant part of the Brillouin zone with substantial occupation probability. 

\section{Correlation function}

We calculate the velocity correlation function $\langle \hat{v}(t) \hat{v}(t+\bar{t})\rangle$ using the quantum regression theorem,  
\begin{equation}
    \langle \hat{v}(t) \hat{v}(t+\bar{t})\rangle = \sum_k \mbox{Tr}_s(\hat{v}_k e^{\mathcal{L}\bar{t}}[\hat{v}_k\hat{\rho}_k(t)]).
\end{equation}
The density matrix $\hat{\rho}_k(t)$ describes the steady state, either with or without momentum relaxation. For short time differences $\bar{t}$ compared to the momentum relaxation time, the correlation function evolves according to the Liouvillian superoperator $\mathcal{L}$ which only incorporates processes due to the dissipator $\mathcal{D}_s$. Writing the density matrix in the vectorized form $\rho = (\langle \tau_{+,k}\tau_{-,k}\rangle,\  \langle \tau_{-,k}\tau_{+,k} \rangle, \ r_{x,k}, \ r_{y,k})^T$, the Liouvillian can be written in matrix form, 
\begin{equation}
\mathcal{L} = \begin{pmatrix} A & 0 \\ 0 & B\end{pmatrix}, \ \ \  A= \sum_{\alpha = h,c}\begin{pmatrix} -\Gamma^+_{\alpha,k} & \Gamma^-_{\alpha,k}\\ \Gamma^+_{\alpha,k} & -\Gamma^-_{\alpha,k}\end{pmatrix}, \ \ \ B = \begin{pmatrix} -\Gamma_k/2 & -2\Delta_k \\ 2\Delta_k & -\Gamma_k/2 \end{pmatrix},
\end{equation}
with $\Gamma_k = \sum_{\ell,\alpha} \Gamma^{(\ell)}_{\alpha,k}$. This follows from the evolution equations given in Eq. (\ref{parameter evolution equations}). The exponential of the Liouvillian reduces to exponentiating the $2\times 2$ matrices $A$ and $B$. For any $2 \times 2$ matrix $M = m_0 + \mathbf{m}\cdot \boldsymbol{\sigma}$, we have  $e^{M\bar{t}} = e^{m_0\bar{t}}(\cosh(m\bar{t}) +\mathbf{m}\cdot \boldsymbol{\sigma}\sinh(m\bar{t})/m) $ with $m = \sqrt{\mathbf{m}\cdot \mathbf{m}}$.  Noticing that
\begin{equation}
\sum_{\alpha}\Gamma^{\pm} _{\alpha,k} = \sum_{\alpha}\frac{\Gamma^{\pm}_{\alpha,k}+\Gamma^{\mp}_{\alpha,k}+\Gamma^{\pm}_{\alpha,k}-\Gamma^{\mp}_{\alpha,k}}{2} = \frac{\Gamma_k}{2}(1\mp r_{z,k}),
\end{equation} 
we find $A = \Gamma_k\left(1+(1,ir_{z,k},r_{z,k})^T\cdot \boldsymbol{\sigma})\right)/2$. Thus, we have
\begin{equation}
    e^{A\bar{t}}  = \frac{1}{2}\left[ \begin{pmatrix} 1+r_{z,k} & 1+r_{z,k}\\ 1-r_{z,k}&1-r_{z,k} \end{pmatrix} +e^{-\Gamma_k \bar{t}}\begin{pmatrix} 1-r_{z,k} & -(1+r_{z,k})\\ -(1-r_{z,k})&1+r_{z,k} \end{pmatrix}\right], \ \ e^{B\bar{t}} = e^{-\Gamma_k \bar{t}/2}( \cos(2\Delta_k \bar{t}) -i\sigma_y \sin(2\Delta_k \bar{t})).
\end{equation} 
In matrix form, we have
\begin{equation}
    \hat{v}_k \hat{\rho}(t) = r_{o,k}\left[ w_o\sin(k) -r_{z,k}\hat{\boldsymbol{\Delta}}_k\cdot \partial_k \mathbf{d}_k +\left(r_{z,k}w_o\sin(k) \hat{\boldsymbol{\Delta}}_k  -\partial_k \tilde{\mathbf{d}}_k\right)\cdot \boldsymbol{\tau}_k \right], 
\end{equation}
which translates into
\begin{align}
    \hat{v}_k \hat{\rho}(t) &= r_{o,k}\left((1+r_{z,k})\left(w_o\sin(k)-\hat{\boldsymbol{\Delta}}_k\cdot \partial_k \mathbf{d}_k\right),(1-r_{z,k})\left(w_o\sin(k)+\hat{\boldsymbol{\Delta}}_k\cdot \partial_k \mathbf{d}_k\right),-\partial_k\tilde{d}_{x,k},-\partial_k\tilde{d}_{y,k}\right)^T \nonumber \\ 
    &= r_{o,k}\left((1+r_{z,k})\partial_k \epsilon_{\uparrow,k},(1-r_{z,k})\partial_k \epsilon_{\downarrow,k},-\partial_k\tilde{d}_{x,k},-\partial_k\tilde{d}_{y,k}\right) = (\boldsymbol{\rho}_{\text{diag}},\boldsymbol{\rho}_{\text{off}})
\end{align}
in the vectorized notation. Here we defined $\tilde{\mathbf{d}}_k$ via $\partial_k\mathbf{d}_k \cdot \boldsymbol{\sigma} = \partial_k\tilde{\mathbf{d}}_k \cdot \boldsymbol{\tau}_k$. Applying the exponentiated Liouvillian to the vectorized $\hat{v}_k \hat{\rho}(t)$, one obtains
\begin{equation}
    e^{A\bar{t}} \boldsymbol{\rho}_{\text{diag}} = r_{o,k}\left[ \frac{(1+r_{z,k})\partial_k\epsilon_{\uparrow,k}+(1-r_{z,k})\partial_k\epsilon_{\downarrow,k}}{2}\cdot \begin{pmatrix} 1 + r_{z,k} \\ 1- r_{z,k}\end{pmatrix}+e^{-\Gamma_k \bar{t}}\frac{(1-r_{z,k}^2)(\partial_k \epsilon_{\uparrow,k}-\partial_k\epsilon_{\downarrow,k})}{2}\begin{pmatrix} 1\\-1 \end{pmatrix}\right],
    \end{equation}
    and
    \begin{equation}
    e^{B\bar{t}}\boldsymbol{\rho}_{\text{off}} = -e^{-\Gamma_k \bar{t}/2}r_{o,k} \begin{pmatrix} \cos(2\Delta_k \bar{t}) \partial_k \tilde{d}_{x,k}-\sin(2\Delta_k \bar{t}) \partial_k \tilde{d}_{y,k} \\ \sin(2\Delta_k \bar{t}) \partial_k \tilde{d}_{x,k}+\cos(2\Delta_k \bar{t}) \partial_k \tilde{d}_{y,k}\end{pmatrix}.
\end{equation}
Reverting to  $2\times 2$-matrix notation, this gives
\begin{align}
    e^{\mathcal{L}\bar{t}}[\hat{v}_k \hat{\rho}(t)] &= r_{o,k}\left\{\left[\frac{(1+r_{z,k})\partial_k\epsilon_{\uparrow,k}+(1-r_{z,k})\partial_k\epsilon_{\downarrow,k}}{2} -e^{-\Gamma_k \bar{t}/2}[\cos(2\Delta_k \bar{t}) \partial_k \tilde{d}_{x,k}-\sin(2\Delta_k \bar{t}) \partial_k \tilde{d}_{y,k}]\tau_{x,k}\right]\right. \nonumber \\ 
    &-e^{-\Gamma_k \bar{t}/2}[\sin(2\Delta_k \bar{t}) \partial_k \tilde{d}_{x,k}+\cos(2\Delta_k \bar{t}) \partial_k \tilde{d}_{y,k}]\tau_{y,k} + \left[ \frac{(1+r_{z,k})\partial_k\epsilon_{\uparrow,k}+(1-r_{z,k})\partial_k\epsilon_{\downarrow,k}}{2} r_{z,k}\right. \nonumber \\
    &\left.\left.-e^{-\Gamma_k \bar{t}} (1-r_{z,k}^2)\hat{\boldsymbol{\Delta}}_k \cdot \partial_k \mathbf{d}_k \right]\tau_{z,k} \right\}.
\end{align}
Finally, we multiply by $\hat{v}_k$ and trace over spin space which then yields the following for the correlation function,
\begin{equation}
    \langle \hat{v}(t) \hat{v}(t+\bar{t})\rangle = \sum_k r_{o,k}\left[\Tr_s(\hat{v}_k(1+r_{z,k}\hat{\boldsymbol{\Delta}}\cdot \boldsymbol{\sigma}))^2 +e^{-\Gamma_k \bar{t}/2}\cos(2\Delta_k \bar{t})(\hat{\boldsymbol{\Delta}}_k \times \partial_k \mathbf{d}_k)^2  +e^{-\Gamma_k \bar{t}}(1-r_{z,k}^2)(\hat{\boldsymbol{\Delta}}_k \cdot \partial_k \mathbf{d}_k)^2\right].
\end{equation}
Noting that
\begin{equation} 
1-r_{z,k}^2 = 4\sum_{\alpha,\alpha'} \frac{\Gamma^+_{\alpha,k}\Gamma^-_{\alpha',k}}{\Gamma_k},
\end{equation}
we obtain the expression given in the main text. We further read off the constant $g_0$ from Eq. \eqref{eq:correlator} to be 
\begin{equation}
    g_0 = \sum_k r_{o,k} \Tr_s(\hat{v}_k(1+r_{z,k}\hat{\boldsymbol{\Delta}}\cdot \boldsymbol{\sigma}))^2 - \langle \hat{v}(t) \rangle\langle\hat{v}(\infty) \rangle.
\end{equation}

As for the average velocity, we look at the limit $T, \Delta_k \ll w_0$. In this limit, the relevant momenta $k$ are close to $k=0$, so that the oscillation period is given by the Zeeman field $\varepsilon$. Moreover, the decay rate is given by $\Gamma_{k=0}=\sum_\alpha \xi_{\alpha,0} \coth(\varepsilon/2T)$. The latter defines the spin relaxation time $t_s$ in this limit. 

\section{Details of numerical calculations} 

We give details on the numerical calculations underlying the results in the main text. We numerically solve 
the Lindblad equation 
\begin{equation}
    \partial_t \hat{\rho} = -i[\mathcal{H},\hat{\rho}] + \mathcal{D}_s[\hat{\rho}] + \mathcal{D}_m[\hat{\rho}] = \mathcal{L}\hat{\rho},
\end{equation}
with $\mathcal{D}_s$ and $\mathcal{D}_m$ being defined in Eqs.\ \eqref{eq: spin dissipator} and \eqref{eq:DmS} respectively. The velocity expectation value is calculated from 
\begin{equation}
    \langle \hat{v}(t) \rangle = \Tr(\hat{v}e^{\mathcal{L}t}\hat{\rho}^{\text{th}}).
\end{equation}
The velocity correlator is obtained using the quantum regression theorem, 
\begin{equation}
    \langle \hat{v}(t) \hat{v}(t+\bar{t}) \rangle = \Tr(\hat{v}e^{\mathcal{L}\bar{t}}[\hat{v}\hat{\rho}(t)]).
\end{equation}
Here, $\hat{\rho}(t)$ may either be the steady state in the intermediate regime, $t_s \ll t \ll t_m$, or the fully relaxed state, $t \gg t_m$. Using $\rho_k = (r_{o,k,\uparrow},r_{o,k,\downarrow},r_{x,k},r_{y,k}$) as components of a $4N$-dimensional vector, the Liouvillian $\mathcal{L}$ becomes a $4N \times 4N$ matrix, 
\begin{equation}
\mathcal{L}\begin{pmatrix}
        \rho_{-\pi} \\ \rho_{-\pi + \delta k} \\ \vdots \\ \vdots \\ \vdots \\ \rho_{\pi - \delta k}\end{pmatrix} = \begin{pmatrix} L_{-\pi} & M_{-\pi, -\pi + \delta k} & M_{-\pi,-\pi+2\delta k} & \dots & \dots & M_{-\pi, \pi - \delta k} \\  M_{-\pi+\delta k,-\pi} & L_{-\pi + \delta k}& M_{-\pi+\delta k,-\pi+2\delta k} & \dots & \dots & M_{-\pi+\delta k, \pi - \delta k} \\ M_{-\pi+2\delta k,-\pi} & M_{-\pi + 2\delta k,-\pi+\delta k}& L_{-\pi+2\delta k} & \dots & \dots & M_{-\pi+2\delta k, \pi - \delta k} \\ \vdots & \vdots & \ddots & \ddots &\ddots &\vdots \\ M_{\pi-2\delta k,-\pi} &  M_{\pi-2\delta k,-\pi+\delta k} & \dots & \dots & L_{\pi-2\delta k} & M_{\pi-2\delta k,\pi-\delta_k} \\ M_{\pi-\delta k,-\pi} &  M_{\pi-\delta k,-\pi+\delta k} & \dots & \dots &  M_{\pi-2\delta k,\pi-\delta_k}  &L_{\pi-\delta k}\end{pmatrix} \begin{pmatrix}
        \rho_{-\pi} \\ \rho_{-\pi + \delta k} \\ \vdots \\ \vdots \\ \vdots \\ \rho_{\pi - \delta k}\end{pmatrix}.
\end{equation}
Here, $\delta k = 2\pi/N$ and $L_{k}$ and $M_{k,k'}$ are $4 \times 4$ matrices, 
\begin{equation}
    L_k = \begin{pmatrix} -l_k^{\uparrow,\uparrow} & l_k^{\uparrow,\downarrow} & & \\ 
    l_k^{\downarrow,\uparrow} & -l_k^{\downarrow,\downarrow} & & \\ & & l_k^{x} & -2\Delta_k \\ & & 2\Delta_k & l_k^{y}\end{pmatrix}, \ \ \ M_{k,k'} = \begin{pmatrix} m_{k,k'}^{\uparrow,\uparrow} & m_{k,k'}^{\uparrow,\downarrow} & & \\ 
    m_{k,k'}^{\downarrow,\uparrow} & m_{k,k'}^{\downarrow,\downarrow} & & \\ & & 0& 0 \\ & & 0 & 0\end{pmatrix}
\end{equation}
with entries
\begin{align}
    l_{k}^{\tau,\tau'} &= \sum_\alpha \xi_{\alpha,k}[\text{n}(2\Delta_k,T_\alpha) + \delta_{\tau',\downarrow}] + \delta_{\tau,\tau'}\sum_{q,\tau''} \frac{\lambda_q}{N} \nu(\Omega_{\tau,\tau''}^{k,q})\left|\braket{\tau_k}{\tau''_{k+q}}\right|^2|\text{n}(\Omega_{\tau,\tau''}^{k,q},T)|, \\
    l_k^{x} &= l_k^{y} =  - \frac{1}{2}\sum_\alpha \xi_{\alpha,k}[2\text{n}(2\Delta_k,T_\alpha) +1] - \frac{1}{2} \sum_{q,\tau,\tau'} \frac{\lambda_q}{N} \nu(\Omega_{\tau,\tau'}^{k,q})\left|\braket{\tau_k}{\tau'_{k+q}}\right|^2|\text{n}(\Omega_{\tau,\tau'}^{k,q},T)|, \\
    m^{\tau,\tau'}_{k,k'} &= \sum_{q} \delta_{q,k-k'}\frac{\lambda_q}{N} \nu(\Omega_{\tau,\tau'}^{k,q})\left|\braket{\tau_k}{\tau'_{k+q}}\right|^2|\text{n}(\Omega_{\tau,\tau'}^{k,q},T)+1|.
\end{align} 

\end{document}